\documentclass[10pt]{article}

\usepackage{newtxtext,newtxmath}

\usepackage{graphicx}
\usepackage{booktabs}
\usepackage{booktabs,graphicx}

\renewenvironment{abstract}
	{\quotation}
	{\endquotation}

\date{}

\makeatletter
\renewcommand{\fnum@figure}{\textbf{Figure \thefigure}}
\renewcommand{\fnum@table}{\textbf{Table \thetable}}
\makeatother

\usepackage{scicite}

\usepackage{url}

\def\scititle{
	Social bots weaken activist cohesion
}
\title{\bfseries \boldmath \scititle}

\author{
	Linda Li$^{1,2\ast}$,
	Orsolya Vasarhelyi$^{3,4}$,
	Balazs Vedres$^{5}$\and
	\small$^{1}$Department of Methodology, London School of Economics\and
    \small$^{2}$Oxford Internet Institute, University of Oxford\and
	\small$^{3}$Center for Collective Learning, Corvinus Institute for Advanced Studies, Corvinus University\and
    \small$^{4}$Institute of Data Analytics and Information Systems, Corvinus University\and
    \small$^{5}$Central European University\and
	\small$^\ast$Corresponding author. Email: l.li88@lse.ac.uk\and
}

\begin{document} 

\maketitle

\begin{abstract} 
Social bots now make up a substantial share of online political communication, where they are studied mainly as producers of misinformation and amplified content. Far less is known about whether their presence reshapes the human relationships that hold movements together. We ask whether exposure to bots during a protest peak is followed by the erosion of cohesion in human networks. Tracking retweet networks of core participants in the 2020 Black Lives Matter (BLM) protests before, during, and after the peak, we measure change in cohesion at two scales: triadic closure in individual ego networks and edge density within detected communities. Greater bot exposure during the peak predicts steeper subsequent declines in human cohesion at both scales, and the loss concentrates among supporters of the movement. Bots may weaken activism less by changing what people believe than by dissolving the ties through which collective action is sustained.
\end{abstract}

\section*{Introduction}

Social bots are now a substantial part of the online public sphere. In political and activist communication, automated or semi-automated accounts can account for a sizeable share of activity: prior research has estimated that algorithmic messaging targeted at activists can reach 20 to 25 percent of communication volume, making exposure to bot-produced content difficult for human users to avoid \cite{Stella2018, yan2018, li2024socialbotssouractivist}. Bots are often studied as sources of misinformation, amplification, or sentiment change in online political communication. We examine a different aspect of automated influence: bots may also be associated with changes in the human network structure of online activism. Civic movements depend not only on the circulation of messages but also on the persistence of human-human ties through which participants recognize one another, coordinate action, form collective identities, and sustain engagement \cite{diani2003networks, diani2000social, bennett2023logic}. If exposure to bots during moments of intense mobilization is followed by weakened human cohesion, then automated activity may affect activism not only by altering discourse but also by disrupting the relational infrastructure that supports collective action.

Online civic networks are vital to contemporary protest. From Occupy Wall Street and the Arab Spring to Black Lives Matter and climate mobilizations, social media platforms have enabled activists to disseminate information, increase visibility, recruit participants, and maintain communities of engagement \cite{Bennett2018, Gonzalez-Bailon2013, Freelon2018, jost2018social, boulianne2020school, Caren2020}. Online social networks are therefore not merely channels for communication. They are social structures in which people find allies, co-interpret events, develop shared understandings, and come to see themselves as part of a collective cause. Although activist communication often occurs in intense bursts, cohesive civic networks can bridge peaks of mobilization by enhancing trust, reducing uncertainty about sustained participation, and reinforcing perceptions of solidarity \cite{tufekci2017twitter}. Cohesion allows activists to identify with durable movement communities between peaks of public attention, rather than dispersing back into the anonymity of social media platforms \cite{gonzalez2011dynamics, Campbell2013, park2015comparing, tindall2004social}.

The same openness that makes online civic networks effective for mobilization also makes them vulnerable to manipulation. During protest peaks, activists, supporters, opponents, media actors, political organizations, and automated accounts may converge on the same hashtags and retweet networks. Prior research has shown that social bots are active in political communication and amplify off-topic messages, increase exposure to inflammatory content, and participate in contentious public conversations \cite{Ferrara2016, shao2018anatomy, Stella2018, howard_lie_2020}. Studies have documented bot activity in elections and referenda, as well as in protest-related communication, including the Yellow Vest movement, anti-corruption protests, climate politics, Black Lives Matter, and protest cases in Latin America \cite{Suarez-Serrato2016, Oliveira2016, Salge2018, Bastos2019, Grinberg2019, Keller2019, Gonzalez-Bailon2021, marlow2021, jones2022out}.

Most research thus far had focused primarily on bots as producers or amplifiers of content, while less is known about whether bot activity is associated with subsequent changes in the cohesion of human-human ties. This distinction is important because network erosion is often non-linear \cite{callaway2000network}: A protest communication network may lose local density and triadic closure (via node and edge loss) while preserving global reachability: users can remain connected through indirect paths even as redundancy declines. But as human-human ties thin out, the network can approach a fragmentation threshold, where small additional losses produce disproportionate changes in component structure, path lengths, and reachability \cite{albert2000error}. Gradual erosion of cohesion may therefore precede abrupt loss of connectivity. Such structural weakening could have important consequences for civic action, because movements depend not only on visibility and message volume but also on resilient patterns of recognition, trust, solidarity, and coordination \cite{diani2003networks, diani2000social, Friedkin2004, tindall2004social, marti2017, Gonzalez-Bailon2023}.

We examine this question in the context of the 2020 Black Lives Matter protests following the death of George Floyd. Black Lives Matter is a relevant case because the movement has relied heavily on social media for visibility, mobilization, and political expression, and previous studies have documented automated and coordinated activity in BLM-related communication \cite{arif2018acting, Freelon2018, jones2022out}. The protests following George Floyd's death generated a large and highly visible wave of online and offline mobilization, providing a setting in which to examine how bot exposure during a protest peak is associated with subsequent changes in human network cohesion \cite{christian2022background}.

We focus on the protest peak from 7 to 8 June 2020 and construct longitudinal communication networks before, during, and after this period. Protest peaks are analytically important because they concentrate attention, interaction, and exposure. They are moments when activists, supporters, opponents, media actors, and bots are most likely to enter the same communication space, making them critical periods for observing how automated activity becomes embedded in human network environments. 

This design allows us to move beyond measuring whether bots were present or active during the peak. Instead, we ask whether human networks with greater bot exposure during the peak experienced larger subsequent declines in cohesion. Network cohesion is not incidental to activism but its structural foundation: the density of human-to-human ties is what lets participants recognise one another as allies, coordinate action, and sustain a shared identity across the lulls between mobilisation waves \cite{diani2003networks, Bennett2018, Gonzalez-Bailon2013}. When this relational scaffolding erodes, a movement can lose the capacity to reassemble even if individual supporters remain present, making cohesion a more fundamental target of disruption than any single message or opinion.

We operationalize human cohesiveness at two network scales. At the ego-network level, we measure changes in the local clustering coefficient of human-only neighborhoods, capturing triadic closure and the persistence of triangles among a user's human contacts from the protest peak to the post-peak period \cite{watts1998}. Ego-network clustering is a broad measure of local closure across users' neighborhoods: it indicates whether the alters around a user remain mutually connected, even when the user remains active in the network \cite{Brooks2014}. At the community level, we detect sub-communities during the protest peak and measure subsequent changes in human-human edge density within those same communities \cite{Moody2003b, palla2005uncovering}. Community density provides a more restrictive meso-scale measure of cohesion, focusing on denser, group-like regions of the network where activists connect into communities. Together, ego-network clustering and community density capture two related but distinct dimensions of human cohesiveness: local triadic closure and group-level cohesion within detected activist communities.

\begin{figure}
\centering
\includegraphics[width=1\linewidth]{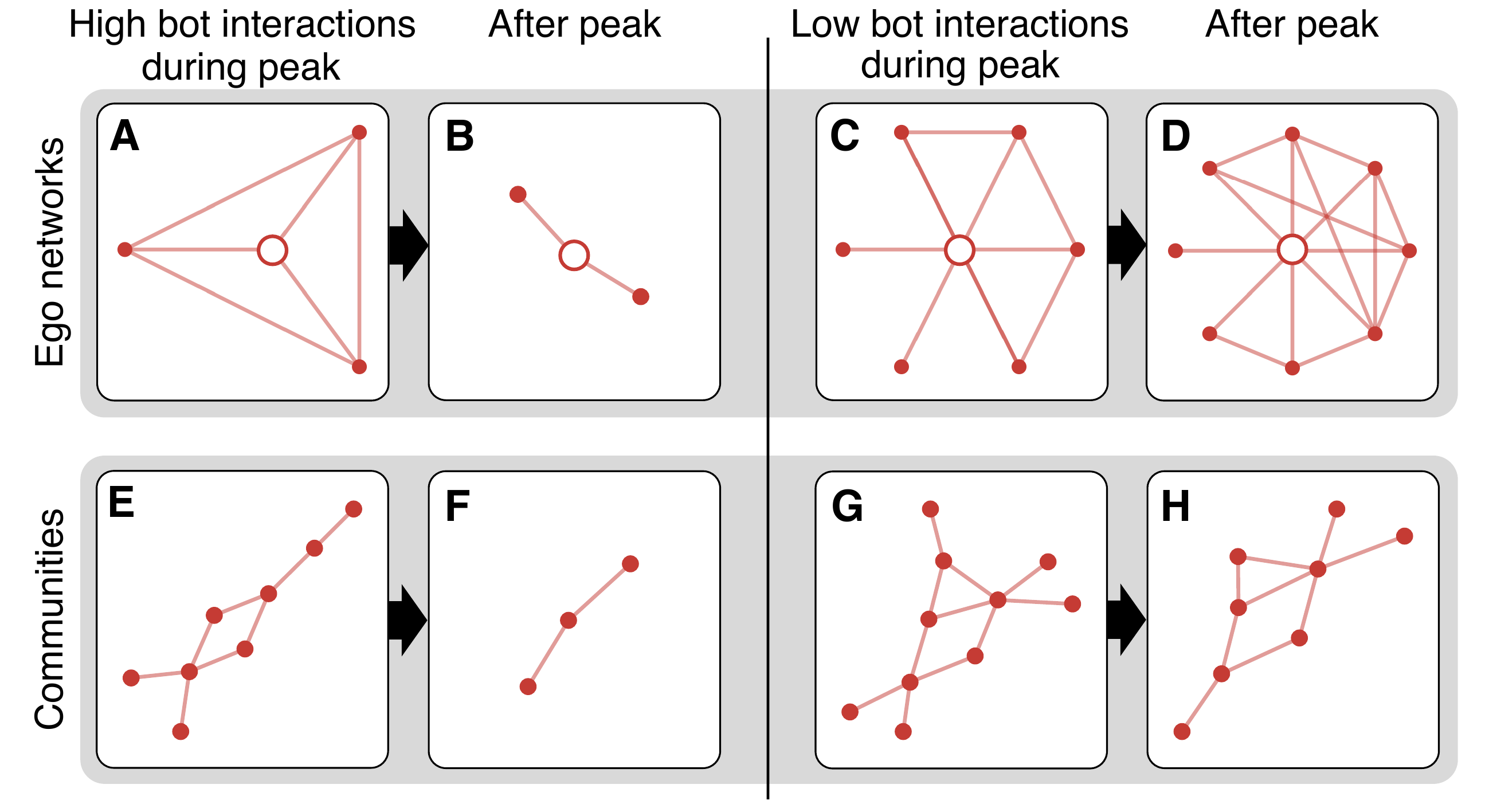}
\caption[Examples of change in cohesion as a function of bot interactions.]{\textbf{Examples of change in cohesion as a function of bot interactions.} Example ego networks (\textbf{A} to \textbf{D}) and communities (\textbf{E} to \textbf{H}) during and after the protest peak, for users with high bot interactions (\textbf{A}, \textbf{B}, \textbf{E}, \textbf{F}) and low bot interactions (\textbf{C}, \textbf{D}, \textbf{G}, \textbf{H}); arrows indicate the during-peak to after-peak transition. Red nodes represent human users; the ego is the large open circle at the centre of each ego network (the ego is excluded from measuring bot interactions and clustering). Clustering coefficients are computed on human-only subgraphs, excluding self-loops, with all edges unweighted.}
\label{fig:network}
\end{figure}

Figure~\ref{fig:network} demonstrates our analytical strategy with our two-level design of ego networks and communities. The figure compares observed examples of ego networks and detected communities, following their network structure from the protest peak to the post-peak period, also distinguishing examples with high and low bot interaction during peak. At the ego network level our focus is on alter-alter dyads, and we track cohesion loss as the disconnection of such alter-alter edges. We record bot interactions during peak for alters only, as we are not recording ego-alter edge disconnection. (Ego networks are overlapping sampling frames to capture clustering change as each node can serve as ego as long as it has at least two edges during peak; we handle the non-independence of observations at the go network level via appropriate generation of standard errors.) 

At the level of communities we start from detection from the peak time period adjacency system, and record community density during peak and after. Cohesion loss at this level appears as density decrease of human-human edges within a given community. We measure bot interactions for all members of a given community, and we use the mean value to characterize a community. (Our units of observation at this level are communities resulting from Clique Percolation Method detection \cite{palla2005uncovering}; thus, node overlaps are permitted. Hence, again, we adopt standard errors suited for non-independent observations.)

We also examine whether the association between bot exposure and cohesion change varies by users' prior orientation toward the movement. Previous research suggests that pre-existing opinions shape how users respond to automated or politically charged communication, and that susceptibility to bot influence may vary across users \cite{Wald2013, Appling2017, Wischnewski2024}. In the context of activism, this heterogeneity is important because supporters, opponents, and neutral observers may occupy different positions in the communication network and may be differently exposed to bot activity. We therefore test whether prior support for Black Lives Matter moderates the relationship between bot exposure and subsequent cohesion change. This analysis allows us to assess whether bot exposure is associated with general weakening across the network or whether the association is concentrated among users more closely aligned with the movement.

We therefore ask a single question: do human networks that were more exposed to bots during the protest peak lose more of their cohesion afterwards? The answer is yes—and consistently so. 

Networks more heavily infiltrated by bots during peak mobilisation show markedly steeper subsequent declines in human cohesion, both in the triadic closure of individual activists' neighbourhoods and in the connective density of whole activist communities. The effect is strongest precisely where movements can least afford it: among the supporters at the network's core. Strikingly, this erosion has little to do with what bots say. Rather than changing minds, automated accounts appear to corrode the relational fabric itself—thinning the human-to-human ties through which activists find one another, coordinate, and hold together between waves of mobilisation. As generative AI makes convincingly human-like bots cheap to deploy at scale, the most serious threat to digital civic life may not be persuasion or disinformation, but the quiet structural dissolution of the communities that make collective action possible.



\section*{Findings}
\subsection*{Hybrid communication network}

\begin{figure}
\centering
\includegraphics[width=1\linewidth]{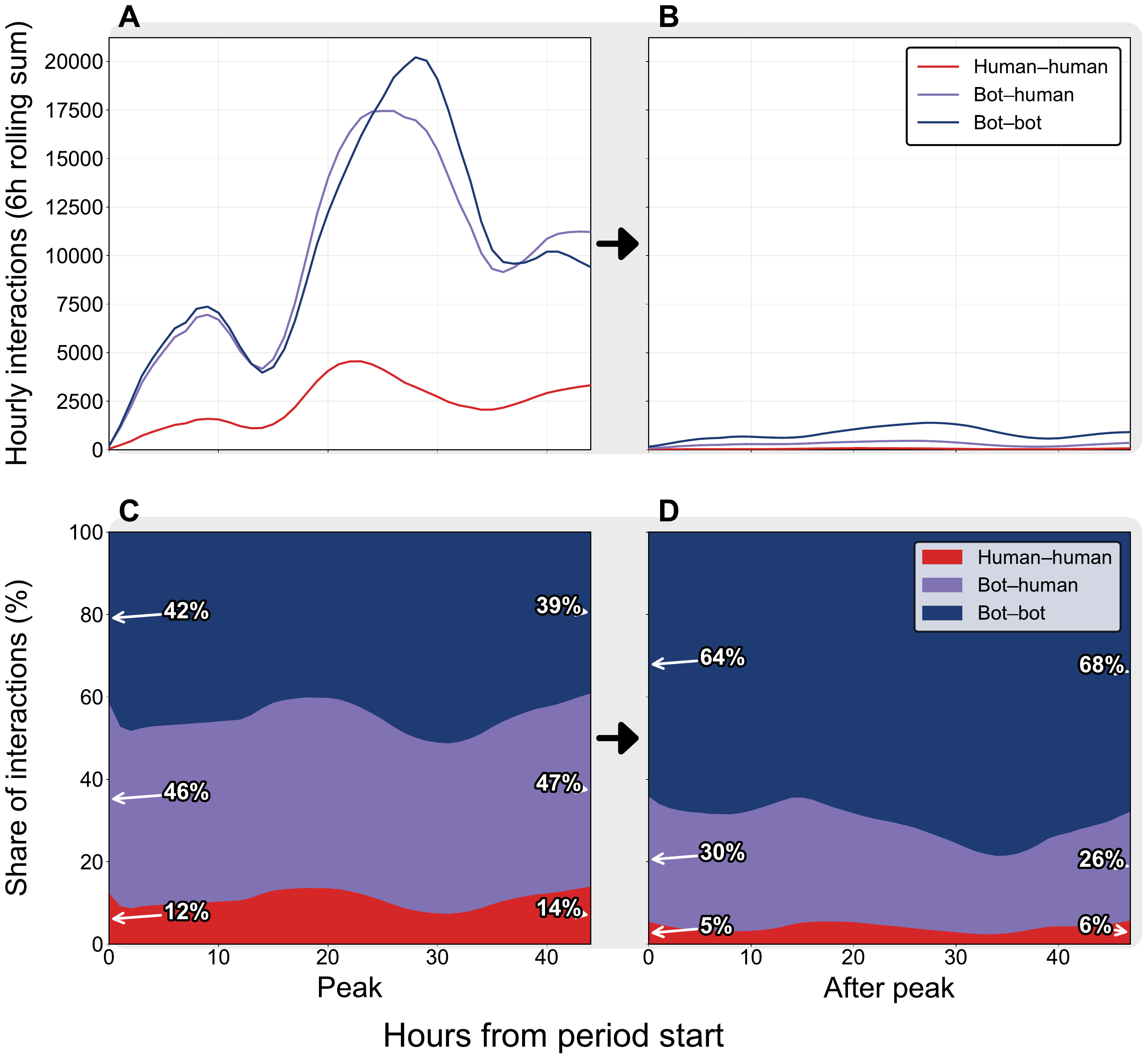}
\caption[Human and bot interaction dynamics on Twitter during and after the George Floyd protests]{\textbf{Human and bot interaction dynamics on Twitter during and after the George Floyd protests.} \textbf{(A, B)} Hourly interaction volume (6-hour rolling sum) during the protest peak \textbf{(A)} and the subsequent protest-after period \textbf{(B)}, separated by interaction type: human--human (red), bot--human (purple), and bot--bot (navy). \textbf{(C, D)} Corresponding share of total interactions (\%) by type over the same two periods, with each type's share labeled at the start and end of each window. The \textit{x}-axis gives hours from the start of each period (protest peak beginning Jun 7--8 UTC, during the George Floyd Black Lives Matter mobilization; protest-after period beginning Sep 1--2 UTC).}
\label{fig:interactions}
\end{figure}

Social bots and human users at the core of the George Floyd protest formed a fragmented but persistent network, which thinned rather than dissolved after the protest peak. We describe this network at the peak and in the post-peak period before estimating how bot exposure relates to changes in human cohesion.

We identified the protest core as the k-core 2 subgraph of the retweet network during the peak, and traced the same users' communication in the post-peak period. At the peak, this core contained 92,078 users (27,221 human, 50,982 bot) and 372,932 edges, of which 55\% (50,982 users) displayed automated-like behavior. (This share is not sensitive to the classification cut-off: it ranges only from 42\% to 55\% across bot-probability thresholds of .65–.75.) The figure is somewhat higher than, but broadly consistent with, the roughly 40\% bot prevalence reported for other large mobilisations such as the Indigenous Rights Movement \cite{gonzalez2011dynamics}.

Figure~\ref{fig:interactions} decomposes interactions by the account types involved. During the peak, bot–bot and bot–human exchanges outnumbered human–human ones in absolute volume (panels A–B), together comprising roughly 88\% of all interactions, against approximately 12\% human–human (panels C–D). In the post-peak period, total volume declined by more than an order of magnitude, but the network did not disappear; its composition shifted further toward automation, with the bot–bot share rising from about 42\% to 68\% and the human–human share falling from about 12\% to 6\%. (Consistent with this, suspected bots were more likely than humans to remain active among the same core users: approximately 69\% of bots versus 60\% of humans still appeared in post-peak retweet interactions overall, and 12\% versus 5\% when restricting to BLM-related retweets.) Note that Figure~\ref{fig:interactions} characterises the protest conversation itself and therefore counts BLM-related interactions in both windows; the network analyses that follow instead use each user's full communication in the post-peak window, so as to measure whether social ties persist as structure rather than whether attention to the topic endures (see Data and methods). We report the BLM-restricted version of the network analyses in the SI, where the estimated associations are unchanged (Table~\ref{tab:ego_blm_after_robust_t065} and Table~\ref{tab:comm_blm_after_robust_t065}).

The human-only structure is sparse and heavy-tailed rather than densely connected. Human–human density is low ($5.6\times10^{-5}$ at peak, $9.0\times10^{-5}$ after), and the network separates into many small, weakly connected clusters with low clustering coefficients (0.029 at peak, 0.030 after). At the ego-network level - our second analytic scale - human-only local clustering is likewise low (mean 0.055 at peak, median 0) and remains modest among egos retained after the peak (mean 0.112, median 0.071; $N=3{,}291$). Repost attention is concentrated on a small set of central actors (unweighted in-degree: median 2.0, mean 6.37, maximum 12,605); this right-skew is consistent with the core–periphery structure reported for online political communication networks \cite{Barbera2015, Bennett2018}. Bot exposure is unevenly distributed throughout the structure: bot–human density in communities ranges from 0 to 1 (median 0.104), so that some clusters of humans are in a largely human environment, while others are heavily interpenetrated by bots. Our analysis exploits this cross-sectional variation in exposure rather than any single temporal trend.

We examine the variation at two scales. At the community level, clique percolation on the peak network yields 677 sub-communities in our analytic sample, each with a human core of at least one human-only triad; 542 of these remain traceable in the post-peak period. At the ego-network level, 3{,}291 neighbourhoods meet our inclusion criteria (egos are humans that are active before, during and after the protest peak). For both, we measure the change in human-only cohesion from peak to post-peak, using each unit's pre-peak state as a baseline control. (Full descriptive statistics per snapshot are reported in Table~\ref{descriptive_stat_network}].)

These descriptives characterize a polarized human-bot hybrid network at the peak: sparse, hub-dominated, unevenly exposed to automation, and only partially retained afterwards. The following sections estimate how bot exposure during the peak is associated with a subsequent change in human–human cohesion.

\subsection*{Bot interaction weakens cohesion}

We first examine the statistical association between bot exposure and cohesion change from peak to the subsequent communication network in a bivariate setting. (Negative values indicate that the human network became less cohesive after the protest peak.) Because cohesion has different meanings at different network scales, we use two measures by design: the change in human-only local clustering for ego networks and the change in human-human edge density for communities.

\begin{figure}
\centering
\includegraphics[width=1\linewidth]{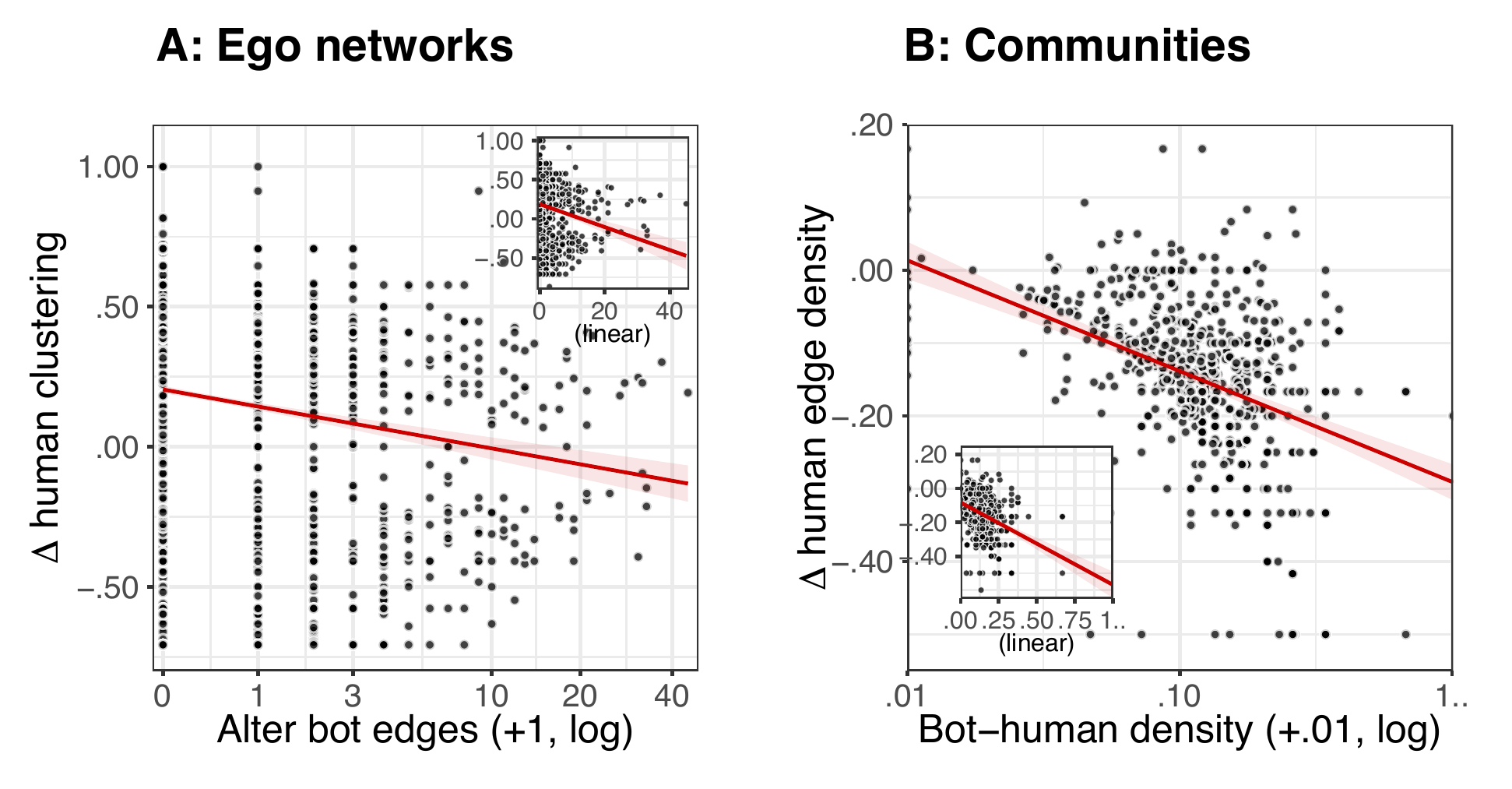}
\caption[Cohesion change versus human--bot interaction.]{\textbf{Cohesion change versus human--bot interaction.} Cohesion is measured by clustering for ego networks and by edge density for communities, two measures chosen by design to suit each structure. (\textbf{A}) For ego networks, change in human clustering (after peak minus during peak) versus the number of edges between a user's alters and bots. (\textbf{B}) For communities, change in human edge density (after peak minus during peak) versus bot--human edge density. Each point is one ego network (\textbf{A}, $n = $~3{,}291) or community (\textbf{B}, $n = $~657); blue lines are linear fits with shaded 95\% confidence intervals. Main panels use a log $x$ axis (offset by 1 in \textbf{A} and 0.01 in \textbf{B}); insets show the linear $x$ axis.}
\label{fig:dot-pred}
\end{figure}

Figure~\ref{fig:dot-pred} reports the association between bot exposure during the protest peak and subsequent changes in human network cohesion at both the ego-network and community levels. Each point represents an ego network (panel A) or a detected community (panel B). Solid lines correspond to linear fits, with shaded regions indicating the associated 95\% confidence intervals.

These two panels capture related but distinct operationalizations of online activist human cohesiveness. Ego-network clustering measures triadic closure in local neighborhoods and therefore includes all closed triads formed among a user's human contacts. Community-level analysis is more restrictive: it considers only closed triads that are embedded within overlapping 4-clique structures identified by clique percolation. As a result, all community-level triangles are also represented in the ego-network analysis, but not all ego-network triangles form part of community structures. The community measure therefore captures cohesion in more tightly integrated, group-like regions of the network. The consistent negative associations across both panels indicate that bot exposure is associated with declines in both general local closure and this more restrictive form of group-level cohesion.

In ego networks (Figure~\ref{fig:dot-pred}A), the horizontal axis measures the number of alter-bot edges observed during the protest peak, while the vertical axis shows the change in human-only local clustering from the peak to the post-peak period. The linear association is negative, higher levels of bot exposure of alters are associated with larger decreases in clustering. The main panel uses a logarithmic scale for the exposure variable (with +1 offset), while the inset presents the same relationship on a linear scale. At the ego network level the intercept is positive, so that without bot interaction there is an increase in clustering from the peak to the period after.  

There is a comparable bivariate association at the community level (Figure~\ref{fig:dot-pred}B). Communities are identified using the Clique Percolation Method with k=4, which detects communities as sets of overlapping four-node cliques (sharing three nodes). Within these communities, bot exposure is operationalized as bot-human edge density during the protest peak, while the outcome variable is the change in human--human edge density between the peak and post-peak periods. The fitted line again shows a negative association, with communities characterized by higher bot-human density during the peak exhibiting larger subsequent reductions in human cohesion. In the community model, the intercept does not differ significantly from zero: net of community size, communities with no bot exposure show no average change in human edge density from peak to post-peak. Bot exposure is associated with declines from this flat baseline.


\subsection*{Greater cohesion loss among supporters}

\begin{figure}
\centering
\includegraphics[width=1\linewidth]{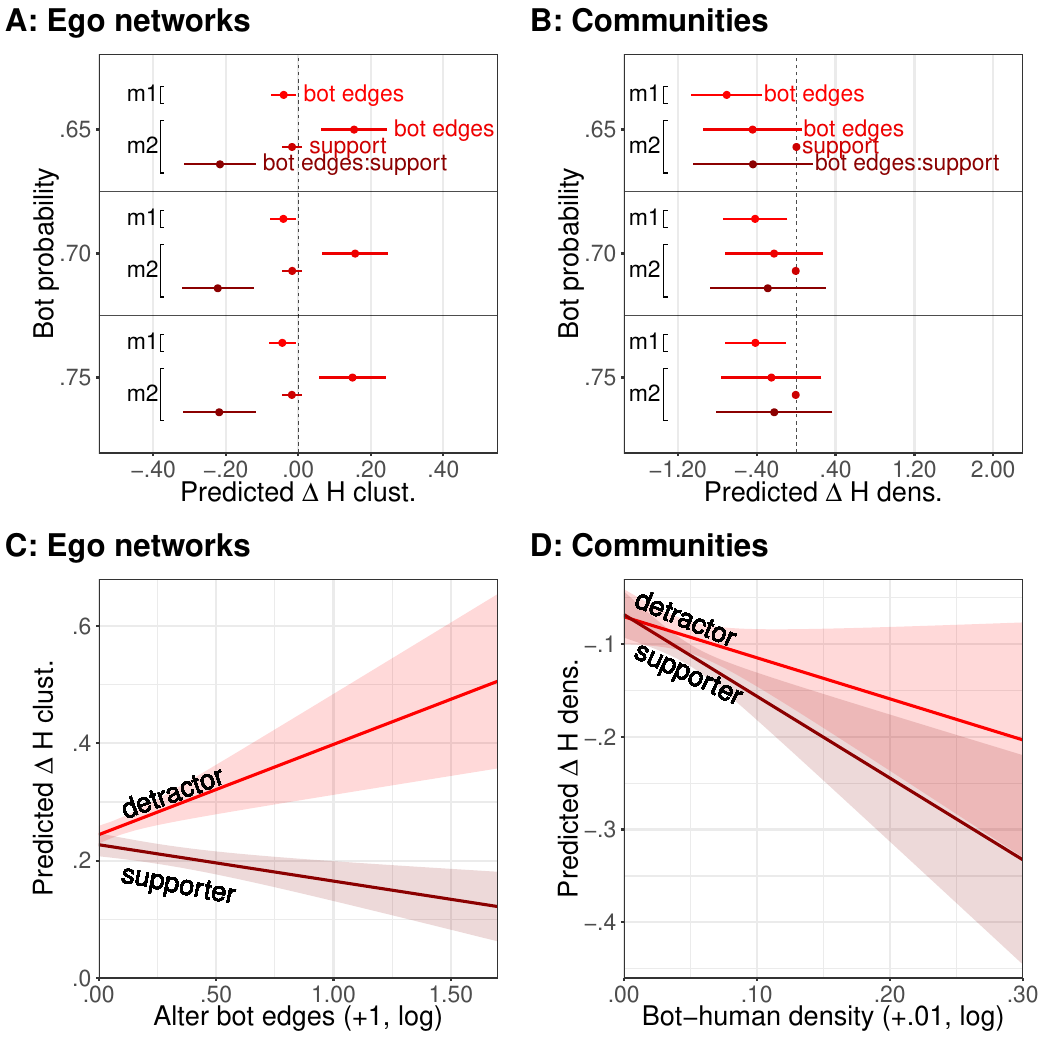}
\caption[Predicted change in cohesion by bot exposure and stance.]{\textbf{Bot exposure erodes predicted cohesion among movement supporters.} (\textbf{A} and \textbf{B}) OLS coefficient estimates with 95\% confidence intervals for the change in human cohesion in ego networks (\textbf{A}, $\Delta$ human clustering) and communities (\textbf{B}, $\Delta$ human edge density), for a model without the bot edges $\times$ support interaction (m1) and one with it (m2), each estimated at bot-probability thresholds of 0.65, 0.70, and 0.75 (rows); the dashed line marks zero. (\textbf{C} and \textbf{D}) Cohesion change predicted by m2 (threshold 0.65) for supporters versus detractors, with 95\% confidence bands: (\textbf{C}) $\Delta$ human clustering versus alter--bot edges, (\textbf{D}) $\Delta$ human edge density versus bot--human density. Supporters have average opinion toward the movement above~0.60. Clustering and edge density are the cohesion measures for ego networks and communities respectively, by design; the $x$ axis in \textbf{C} and \textbf{D} is log-scaled (offset by 1 and 0.01).}
\label{fig:line-pred}
\end{figure}

After identifying a consistent negative bivariate relationship, we next add control variables and examine whether the association between bot exposure and cohesion change varies by users' prior orientation toward the protest movement. Figure~\ref{fig:line-pred} summarizes the results of regression models estimated at both the ego-network and community levels, along with the predicted values from interaction models.

Panels A and B report coefficient estimates from ordinary least squares models predicting change in human cohesion. For each outcome, we estimate two specifications: a baseline model (m1) including bot exposure and controls, and an interaction model (m2) that additionally includes the interaction between bot exposure and prior support for Black Lives Matter. Prior support is derived from users’ tweet histories using a large language model (ChatGPT 3.5), which classifies stance toward the movement on a continuous scale from -1 (strong opposition) to +1 (strong support). For the regression analysis, this variable is operationalized as a binary indicator, coded as 1 if the user’s pre-peak support level exceeds the sample median and 0 otherwise. At the community level, this measure is aggregated across users to characterize the overall orientation of each community. Estimates are shown for three bot-probability thresholds (0.65, 0.70, and 0.75). Points indicate coefficient estimates and horizontal bars represent 95\% confidence intervals. 

At the ego-network level (Figure~\ref{fig:line-pred}A), the coefficient for alter--bot edges is negative across specifications, consistent with the bivariate pattern observed in Figure~\ref{fig:dot-pred}A. When the interaction term is included (m2), the coefficient for the interaction between bot exposure and prior support is also negative, indicating that the decline in clustering associated with bot exposure is more pronounced among users with higher prior support for the movement. The confidence intervals for these interaction terms exclude zero for the main threshold specification, suggesting that the moderating effect is not driven by sampling variability.

A similar pattern appears at the community level (Figure~\ref{fig:line-pred}B). The coefficient for bot--human density is negative across specifications, replicating the bivariate association in Figure~\ref{fig:dot-pred}B. The interaction between bot exposure and average community support is also negative, indicating that communities with stronger pro-movement orientation exhibit larger declines in human--human edge density when bot exposure is higher. The consistency of the negative interaction across threshold specifications suggests that this pattern is robust to alternative definitions of bot accounts.

Panels C and D visualize these interaction effects using predicted values from the m2 model at the 0.65 bot-probability threshold. In ego networks (Figure~\ref{fig:line-pred}C), predicted changes in human clustering are plotted against alter--bot edges separately for supporters and detractors. Among supporters, predicted cohesion declines as bot exposure increases, showing a clear negative slope. Among detractors, the relationship is substantially flatter, with predicted cohesion remaining closer to zero across the observed range of exposure. This divergence indicates that the association identified in the bivariate analysis is largely driven by users aligned with the movement.

At the community level (Figure~\ref{fig:line-pred}D), the same pattern is observed. Communities composed predominantly of supporters show a pronounced negative relationship between bot--human density and predicted cohesion change, while communities with lower or opposing orientation exhibit weaker or near-zero associations. The predicted lines and their confidence bands indicate that the reduction in cohesion is concentrated in communities where movement support is higher.

Across all panels, the direction and relative magnitude of the estimated effects remain stable across bot-probability thresholds. This stability indicates that the moderating pattern does not depend on a specific classification cutoff used to identify bots. Instead, the results consistently extend the bivariate findings by showing that the negative association between bot exposure and cohesion loss is conditional on prior support for the movement.

(See SI Model Table Table~\ref{tab:ego_t065} to Table~\ref{tab:ego_t075} for ego-level models, and SI Model Table~\ref{tab:comm_t065} to Table~\ref{tab:comm_t075} for community-level models.)

Taken together, Figure~\ref{fig:line-pred} complements the descriptive patterns in Figure~\ref{fig:dot-pred} in two ways. First, it confirms that the negative association between bot exposure and human cohesion persists after accounting for baseline structure and user characteristics. Second, it shows that this association is not uniform across the network, but is concentrated among users and communities that are more supportive of the movement. At the ego level, this appears as a greater loss of triadic closure in the neighborhoods of supportive users. At the community level, it appears as a stronger decline in human--human density within communities organized around movement-aligned participants.

\section*{Discussions}

The presence of automated agents in online political discourse has reached a scale that fundamentally alters the nature of the digital public sphere \cite{Stella2018, yan2018, li2024socialbotssouractivist}. While extensive scholarship has documented how social bots manipulate sentiment, spread misinformation, and amplify politically divisive content \cite{Stella2018, Ferrara2016, shao2018anatomy, howard_lie_2020}, the downstream consequences of this automated activity on the relational infrastructure of human-to-human ties have remained largely overlooked. Our findings reveal a critical, structurally disruptive dimension of automated influence: the systematic erosion of the human web that sustains collective action \cite{diani2003networks, diani2000social, Bennett2018}. By tracking the longitudinal retweet networks of core participants in the 2020 BLM protests, we demonstrate that exposure to social bots during peak mobilization is associated with a subsequent decay in human network cohesion that persists months after the protest has subsided. This degradation occurs simultaneously across two distinct topological scales. At the local scale, human ego networks with greater bot exposure exhibit a pronounced loss of triadic closure. At the meso scale, human communities heavily interpenetrated by bots during the protest peak suffer the largest subsequent drops in human edge density. Together, these results indicate that automated activity does not merely alter the informational flow of online mobilization; it actively disrupts the local social configurations through which activists recognize allies, coordinate collective behavior, and maintain shared identities between mobilization waves \cite{Bennett2018, Freelon2018, Gonzalez-Bailon2013, jost2018social, Caren2020, tufekci2017twitter}.

The erosion of cohesion has severe potential implications for the resilience of social movements. By targeting and degrading local closure, social bots facilitate a structural displacement, that operates not by persuading human users to adopt automated opinions, but by introducing noise that renders human-to-human communication less socially rewarding \cite{gonzalez2011dynamics, Campbell2013, park2015comparing, tindall2004social, Gonzalez-Bailon2023}. Our analysis shows that this decay is structurally asymmetric, concentrating mostly among users and communities supportive of the movement. This finding aligns with evidence that pre-existing political orientations condition how individuals engage with politically charged automation \cite{Wald2013, Appling2017, Wischnewski2024}. Because supporters tend to occupy the network core \cite{Barbera2015}, they are disproportionately targeted by automated accounts attempting to amplify, distract, or disrupt discourse \cite{arif2018acting}. This asymmetric disruption breaks down local solidarity where it is most needed, rendering activist networks vulnerable to sudden collapse once public attention recedes \cite{tufekci2017twitter}.

These insights necessitate a conceptual shift in how researchers, platform operators, and policymakers evaluate the threat of automated accounts in democratic spaces. Traditional mitigation efforts focus primarily on informational hygiene, tracking the diffusion of disinformation \cite{howard_lie_2020}. Our study suggests that such approaches overlook the structural impact of bot penetration. To safeguard online civic life, platforms should also track bot-human relational density in communication networks, and the degradation of human ties. This is especially acute now, as the proliferation of generative artificial intelligence lowers the barrier for deploying automated accounts that mimic human interaction with exceptional context-sensitivity. The primary threat of highly sophisticated automation may not lie in its power of ideological persuasion, but in its capacity to occupy online social spaces in ways that fragment human participation to the extent that movements will disappear. Preserving the integrity of digital activism therefore requires protecting not only what people see, but also the relational scaffolding that makes collective action possible.

\section*{Data and methods}

\subsection*{Data}
This research examines the human and bot networks surrounding the Black Lives Matter (BLM) protests on X (formerly Twitter), focusing on discourse following George Floyd's death on 25 May 2020. We defined 7–8 June 2020 as the peak window of protest discourse, selected to capture Twitter activity immediately following the largest day of offline mobilisation. On Saturday 6 June, the Crowd Counting Consortium recorded more than 700 protest events across over 600 U.S. localities \cite{countingcrowds2021}, the single largest day of the George Floyd uprising; we collected all BLM-related tweets over the 48-hour window spanning the peak of this mobilisation. Because online protest discourse typically crests alongside and just after offline events, this window captures the movement at its point of maximum mobilisation \cite{steinertthrelkeld2015online, nicoletti2022tweets}.

To capture relevant discourse, we collected all tweets posted during this period containing BLM-related keywords (e.g. `BLM', `George Floyd', and related terms, including common spelling variants; see SI Table~\ref{tweet-collection-keywords} for the full list). Data were gathered through the Twitter Academic Product track API, which provides a comprehensive archive of historical tweets. Alongside tweet content, we collected metadata including user IDs, which we used to retrieve further account information via the Twitter user-lookup API. The final dataset comprises approximately 2.5 million tweets from more than one million users, posted between 7 June 2020 00:00 and 8 June 2020 00:00.

\subsection*{Bot detection}
We adopted and fine-tuned a bot detection method which involves cross-validation and solid robustness checks developed by our past research\cite{li2024socialbotssouractivist}. This involved using both the publicly available tool, Botometer\cite{yang2022botometer}, and a self-trained bot detection algorithm. `Social bots' were defined as users with a Botometer score threshold above 0.65 and identified as bots by the trained classifier. Accounts scoring between 0.5 and 0.65 were labelled as `unknown', and the rest were classified as `humans'. In the analyses that follow, we assess the sensitivity of our results to this cut-off by re-estimating all models across bot-probability thresholds of 0.65, 0.70, and 0.75, (See SI bot detection for the full procedure)

\subsection{Network construction and filtering}

We constructed a directed, unweighted retweet network (RT network) in which users are nodes and retweets or quotes between users are edges. Each edge records only the presence of a connection: multiple interactions between the same pair of users are collapsed into a single edge. This aggregation is appropriate to our question, which concerns whether interaction with bots affects the connections between human users rather than the intensity of any single tie, and follows established practice in research on online communication networks \cite{Tantardini2019, sen2016focal}, particularly in political communication \cite{valenzuela2020ties}. Here, retweets are posts shared without added comment, and quotes include additional commentary; each is recorded as a directed edge from the original poster to the retweeter or quoter. Self-loops were retained for users who retweeted or quoted their own posts or posted original tweets, so that content origination — not only relaying — is represented in the network. Users with no original tweets or retweets were excluded.

Because online protest networks typically exhibit a hierarchical core–periphery structure \cite{Barbera2015}, we restricted the peak network to its core, following prior work on network cohesion \cite{Jasny2023, yan2018}. 

We used the k-core metric, which identifies the maximal subgraph in which every node has at least k connections \cite{peng2014}, retaining users with a k-core value of at least two. This yielded a network of 92,078 users and 372,932 edges (computed with NetworkX \cite{hagberg2008}). Cohesion measures reported below are computed on undirected projections of these networks, so that density and clustering are defined consistently across snapshots.
To measure change in cohesion, we constructed a post-peak snapshot for the same core users, covering 1–2 September 2020, and a pre-peak baseline snapshot covering 1–2 May 2020. The pre-peak snapshot is used solely to construct baseline control variables and does not enter as a modelled endpoint; the analysis outcome is the change in human-only cohesion from the peak to the post-peak period. Because the pre-peak window precedes George Floyd's death on 25 May 2020, it captures these users' baseline communication structure prior to the mobilisation, providing a pre-protest reference point against which peak and post-peak cohesion can be interpreted. 

Whereas the peak network is restricted to BLM-related communication (see Data), the pre- and post-peak snapshots use users' full-timeline communication rather than BLM-keyword-filtered edges. This asymmetry is deliberate: our aim in the before and after snapshots is to test whether the mobilised core persists as a social structure, not whether attention to the protest topic endures. Because topical attention decays over time, restricting these edges to BLM content would conflate network persistence with topic decay; using full communication instead reveals whether ties among core users survive once the specific topic cools.

As a robustness check, we reconstructed the pre- and post-peak snapshots using BLM-related edges only — matching the edge definition of the peak network — and recomputed every predictor and outcome on the restricted networks; the results are substantively unchanged (see SI Table~\ref{tab:ego_blm_after_robust_t065} and Table~\ref{tab:comm_blm_after_robust_t065}), confirming that our conclusions do not depend on the broader edge definition.

\subsection*{Opinion categorisation using ChatGPT}

To capture users' prior orientation toward the movement, we classified each user's stance on the Black Lives Matter protests, motivated by evidence that pre-existing opinions shape how users respond to bot interaction \cite{Wischnewski2024}. Stance was scored on a continuous scale from -1 (strong opposition) to 1 (strong support), with 0 denoting neutrality or unrelated content, derived from each user's tweets in the relevant snapshot. For the moderation analysis we dichotomised this score at the sample median (0.6) into `Supporters' (pro-protest) and `Detractors' (neutral or anti-protest); neutral and opposing users are grouped together because our theoretical contrast is between movement supporters and all other users. We used each user's pre-peak stance as the prior-opinion measure.

Because the corpus is too large to hand-code, classifications were produced with OpenAI's GPT-3.5 \cite{chatgpt}, an approach validated for stance and content classification in prior work \cite{zhu2023can, wu2023event, gilardi2023chatgpt}. Prompts were developed through two rounds of human-in-the-loop refinement: beginning from a pilot of 100 tweets, two coders reviewed GPT-3.5's output and documented systematic errors, after which prompts were revised with explicit definitions and examples targeting each error type. 

We validated the classification against manual coding on a random sample of protest users: GPT-3.5's support scores correlated with the averaged judgement of two independent coders at r = 0.88, exceeding the correlation between the two coders themselves (r = 0.71), indicating that the model agrees with the human consensus at least as strongly as the coders agree with one another. This validation was conducted on Extinction Rebellion protest data using the same movement-agnostic prompt instrument applied here; because the instrument classifies support for or opposition to the relevant movement rather than any BLM-specific content, we expect its validity to transfer to the present setting. Full validation details (Table~\ref{chatgpt-pearson} are reported in the SI Opinion categorization using ChatGPT.

\subsection*{Community detection}
Because our primary interest lies in how social bots reshape social cohesiveness at the collective level, community-level analysis is the main analytical focus of this study, with individual-level analysis serving as a complementary perspective. We treat sub-communities formed during the protest peak as the fundamental units of analysis, following established practice in network research on meso-level structural change during heightened mobilisation \cite{gonzalez2016networked}. This lets us capture changes in cohesion that are not reducible to individual behaviour but emerge from patterned interactions among groups of users.

We detected sub-communities in the peak k-core-2 network using the Clique Percolation Method (CPM). CPM identifies communities as sets of adjacent k-cliques, where two k-cliques are adjacent if they share $k-1$ nodes, yielding overlapping, densely connected sub-graphs rather than a disjoint partition. This makes CPM well suited to social interaction networks characterised by local density and structural overlap, for two reasons. First, our question concerns human–human cohesion within the network, and CPM is designed to recover small, dense components embedded in large sparse graphs -- precisely where cohesive human interaction is most likely to occur -- isolating tightly knit groups even when the wider network is dominated by weak or transient ties \cite{palla2005uncovering}. Second, its local, clique-based construction avoids repeated global optimisation, making it computationally efficient at the scale of our network \cite{kumpula2008sequential}.

Using CPM with clique size k=4, we identified 4,042 sub-communities containing both bot and human nodes. To ensure each retained community represented meaningful human social structure, we excluded communities with fewer than three human nodes; this guarantees at least one human triad within each community and thus non-trivial human–human interaction. After this constraint, 642 sub-communities remained for analysis.

To examine change in community-level cohesion, we measured human–human edge density within each retained community at the protest peak and in the post-peak period, using the same node set throughout and computing all measures on undirected projections. The pre-peak snapshot was used only to construct baseline controls, not as a modelled endpoint; the outcome is the change in human–human density from peak to post-peak. Tracking identical communities across periods isolates change in internal cohesion from change in community composition, allowing direct assessment of how meso-level human cohesion relates to bot presence during the peak.

\subsection*{Statistical analysis design}
We used two complementary models to assess how bot interaction relates to network cohesion in online activism: an individual-based (ego-network) model and a community-based model. Separating them is warranted methodologically, because each uses a distinct sampling strategy and captures a different structural property of the network, and substantively, because convergent results across two representations of social structure strengthen the empirical claim. 

The two models differ in sampling by design: the individual model restricts attention to users active across the pre-peak, peak, and post-peak periods, enabling assessment of change in individual connectivity, whereas the community model captures collective interaction within broader structures and may include users appearing only in specific phases. Using both reduces bias from relying on any single network representation.

\subsubsection*{Community network analysis}
We measured change in human–human cohesion at the community level as the change in edge density of the human-only subgraph between the peak and post-peak snapshots (computed on undirected projections). Network density is the proportion of observed ties to possible ties among a set of nodes \cite{wasserman1994social}; human–human density therefore captures how extensively human users are connected relative to the maximum possible, making it a size-normalised measure of collective cohesion suited to community-level comparison \cite{Friedkin2004, gonzalez2011dynamics}. We adopt it as our primary outcome.
To quantify bot interaction at the peak, we computed bot–human density within each community, defined as observed bot–human ties over all possible bot–human ties:

\begin{equation}
\text{Bot–human density} = \frac{N_{\text{bot–human edges}}}{N_{\text{human}} \times N_{\text{bots}}}
\end{equation}

Normalising by the number of potential bot–human connections ensures that community size does not mechanically drive the estimated association \cite{borgatti1997network}.

We estimated the relationship between peak bot–human density and the change in human–human density (peak to post-peak) using ordinary least squares (OLS). Alongside the key predictor, we included network-level controls — pre-peak human–human density (baseline) and the number of human nodes — and individual-level controls aggregated to the community level. The latter follow evidence that a user's friend-to-follower ratio shape their role in protest information diffusion \cite{Gonzalez-Bailon2013}. We omit the received-to-sent message ratio, as it is strongly collinear with the network-structural measures that constitute our outcome and controls. Other controls include account age, number of statuses, and number of favourites, as they are measures of tenure and activities relevant to cohesion \cite{kim2023}. Each individual-level control was aggregated across a community's human users by mean or median depending on its distribution (see SI).

\subsubsection*{Individual network analysis}
We estimated the same relationship at the individual level. The sample comprised users with a k-core value of at least two who were active across the pre-peak, peak, and post-peak periods; we excluded nodes isolated in the human-only k-core-2 peak network, since these had no human–human connections and entered only through bot interaction. For each remaining user we extracted their ego network.

We operationalised bot interaction as the number of edges connecting the ego's alters to bots during the peak, excluding edges incident to the ego itself (i.e. the count of alter–bot edges). We use alter–bot rather than ego–bot edges for two reasons. First, alter–bot edges capture how far automated accounts have penetrated the ego's local environment, rather than the ego's own choice to engage with bots — matching our interest in how bot activity reshapes the structural conditions around a user rather than individual engagement, which is subject to self-selection. Second, ego-incident edges partly constitute the ego network's structure, so including them would create a mechanical relationship with the cohesion outcome; restricting to alter–bot edges keeps predictor and outcome analytically separate.

Our outcome was the change in the ego network's local clustering coefficient (human-only, undirected) between the peak and post-peak periods. The local clustering coefficient measures the proportion of a node's neighbours that are themselves connected \cite{watts1998}, an established indicator of structural cohesion \cite{kartun2019}. For node $i$ it is

\begin{equation}
 C_i = \frac{2e_i}{k_i(k_i - 1)},
 \end{equation}

where $e_i$ is the number of edges between node $i$'s neighbours and $k_i$ is the degree of node $i$.
 
We estimated the association using OLS, controlling for baseline network position (the ego's degree and clustering in the pre-peak human-only network) and the same user-level characteristics as above (friend-to-follower ratio, account age, statuses, favourites), each aggregated across the ego's human alters by mean or median (see SI).

\subsubsection*{Opinion dynamics}

To test whether prior opinion moderates the association between bot interaction and cohesion change, we included users' pre-peak support for the movement, motivated by evidence that prior opinions shape susceptibility to bot influence \cite{Wischnewski2024} and that bots may disproportionately target particular viewpoints \cite{varol2017online}. This is a binary indicator (1 if pre-peak stance exceeds the sample median, 0 otherwise). Interacted with bot exposure, it tests whether the association between local bot infiltration and cohesion change differs between supporters and other users.

\subsection*{Robustness}
Our conclusions are stable across several checks; full specifications and coefficients are reported in SI Model Tables (Table~\ref{tab:ego_t065} to Table~\ref{tab:ego_network_error_t065}.)

\paragraph{Bot-classification threshold.}The estimated associations hold across bot-probability thresholds of 0.65--0.75, indicating they are not an artefact of the bot/human cut-off. We do not test thresholds below 0.65 because, in this setting, false positives are a greater concern than false negatives: misclassifying a human as a bot would spuriously inflate estimated bot exposure and bias the association we seek to measure, whereas failing to detect some genuine bots is more conservative. Restricting classification to higher-probability thresholds therefore guards against overstating bot effects.

\paragraph{Edge definition.}To address the asymmetry between the BLM-filtered peak network and the full-timeline pre- and post-peak networks, we reconstructed the pre- and post-peak snapshots using BLM-related edges only---matching the peak's edge definition---and recomputed every predictor and outcome. The estimates are substantively unchanged.

\paragraph{Network autocorrelation.}Because overlapping ego networks and CPM communities share nodes, observations may not be independent. We assessed residual autocorrelation using Moran's~$I$. At the community level, residuals showed no significant autocorrelation in either the main-effect or the opinion-interaction model (all $p \geq 0.67$), so we retain the OLS specification. At the individual level, residuals were clearly dependent: under ego-neighbourhood overlap weights, Moran's~$I = 0.147$ ($p = 0.0001$) for the main-effect model and $I = 0.144$ ($p = 0.0001$) for the interaction model, corroborated under CPM co-membership weights ($I \approx 0.089$, $p < 0.001$).

We therefore re-estimated the individual-level models as maximum-likelihood network error models with row-standardized weights, using two alternative weight matrices: (i)~ego-neighbourhood overlap, defined as the Jaccard similarity between closed human ego neighbourhoods (retaining each ego's top 20 neighbours before symmetrization); and (ii)~CPM co-membership, connecting egos assigned to the same peak community. The network error parameter is significant throughout ($\lambda \approx 0.34$ under overlap weights, $\lambda \approx 0.075$ under co-membership weights; all $p < 0.001$), and filtered residual Moran tests become non-significant (e.g.\ $I = -0.02$, $p \geq 0.078$), confirming that the dependence is removed. The bot-exposure associations survive the correction under both weight matrices: in the main-effect model, peak bot exposure remains negatively associated with the change in human clustering ($\beta = -0.040$, $\mathrm{SE} = 0.019$); in the opinion-interaction model, the positive main term is retained ($\beta = 0.125$, $\mathrm{SE} = 0.045$) and the bot-exposure $\times$ pro-BLM interaction remains negative and strongly significant ($\beta = -0.190$, $\mathrm{SE} = 0.050$). Model fit is essentially unchanged from OLS (pseudo-$R^2 \approx 0.69$; $N = 3{,}291$).


\clearpage 

%

\renewcommand\refname{References and Notes}
\bibliography{references} 
\bibliographystyle{sciencemag}

%
%
%
%
%
%


\section*{Acknowledgments}

\paragraph*{Funding:}
OV was funded by the European Union under Horizon EU project LearnData, grant number: 101086712.

\paragraph*{Author contributions:}
Conceptualization: LL,BV. Data curation: LL. Methodology: LL,BV. Formal analysis: LL,OV,BV. Validation: OV,BV, Investigation: LL. Visualization: LL,OV,BV. Writing—original draft: LL. Writing—review and editing: LL,OV,BV. Supervision: BV.

\paragraph*{Competing interests:}
The authors declare that they have no competing interests.

\paragraph*{Data and materials availability:}
All data needed to evaluate the conclusions in the paper are present in the paper and/or the Supplementary Materials. In accordance with the X (formerly Twitter) Developer Agreement and Policy, which prohibits redistribution of hydrated tweet content, we cannot publicly share the raw tweets or user profiles. The tweet IDs of all collected posts, the derived network files (node and edge lists with bot/human labels and opinion scores), and all analysis code required to reproduce our results are available in a private repository for review (https://osf.io/akp8f/overview?view\_only=5d5a8d6019f946bdad165b8fbd74f539) and will be deposited in a public repository with a permanent DOI upon acceptance. Hydrated tweet content can be reconstructed from the tweet IDs by researchers with valid X API access.


\subsection*{Supplementary materials}
Data collection\\
Bot Identification\\
Opinion categorization using ChatGPT\\
Network constructing and filtering\\
Tables S1 to S12\\
References \textit{(7-\arabic{enumiv})}\\ 
Movie S1\\
Data S1


\newpage


\renewcommand{\thefigure}{S\arabic{figure}}
\renewcommand{\thetable}{S\arabic{table}}
\renewcommand{\theequation}{S\arabic{equation}}
\renewcommand{\thepage}{S\arabic{page}}
\setcounter{figure}{0}
\setcounter{table}{0}
\setcounter{equation}{0}
\setcounter{page}{1} 


\begin{center}
\section*{Supplementary Materials for\\ \scititle}

Linda Li $\ast$,
Orsoloya Vasarhelyi,
Balazs Vedres\\ 
\small$^\ast$Corresponding author. Email: l.li88@lse.ac.uk\\
\end{center}

\subsubsection*{This PDF file includes:}
Data collection\\
Bot Identification\\
Opinion categorization using ChatGPT\\
Network constructing and filtering\\
Tables S1 to S12\\

\newpage


\section*{Data collection}
For our investigation into human and bot networks related to the Black Lives Matter (BLM) protests on X (formerly Twitter), we concentrated on the period from June 7 to 8 June, 2020. This interval was identified as the peak of protest-related activity following George Floyd’s death on 25 May, 2020.

We made use of the Twitter Academic Product track API to collect tweets from this period, allowing us to obtain a comprehensive archive of tweets containing keywords such as `BLM,' `George Floyd,' and several variations with minor spelling differences. The full list of keywords employed is provided in Table \ref{tweet-collection-keywords}.

\begin{table}[!ht]
\centering
\caption{Keywords for tweet collection.}
\begin{tabular}{@{}l@{}}
\toprule
\textbf{Keywords}      \\ \midrule
BlackLivesMatter             \\
black lives matter           \\
BLM  \\
George Floyd  \\
georgefloyd                     \\
 \bottomrule
\end{tabular}
\label{tweet-collection-keywords}

\end{table}

Along with tweet content, we collected associated metadata, including user IDs. We then used these IDs to fetch additional user information via the Twitter user lookup API, which was available by the time of data collection. Our final dataset includes over 2,500,000 tweets from more than 1,000,000 users within the specified time frame, from June 6, 2020, 00:00 to June 8, 2020, 23:59.

Given that the API provided a complete archive of tweets, we consider our sample to be relatively comprehensive for capturing all relevant communication related to the Black Lives Matter (BLM) protests on X (formerly Twitter). However, it is important to acknowledge that the data collection was conducted retrospectively. Twitter's moderation policies, which include censorship of spam messages, and the potential for users to delete their accounts or posts between the protests and the time of data collection, may affect the completeness of the dataset. Additionally, due to the contentious nature of the protests and the often ephemeral behaviour of social bots, there is an increased likelihood that bot-generated messages were deleted or accounts suspended, which could lead to an underestimation of bot activity compared to human users.

Nevertheless, the full Academic Track API has been discontinued\cite{verge_twitter_musk}, adding significant value to our dataset. With the API no longer available, our dataset represents a unique and comprehensive snapshot of protest-related discourse during a critical period. This discontinuation makes our data particularly valuable for historical and comparative analyses, as it captures a complete view of social media activity that is no longer accessible through current tools.

\section*{Bot Identification}
Bot identification follows a methodology developed in prior work \cite{li2024socialbotssouractivist}. That approach uses the Botometer tool, which calculates bot probabilities from a range of features related to user profiles and activity, and applies thresholds (0.65 and 0.5) to classify accounts as `automated', `human', or `unknown'. In addition to Botometer, it trains several supervised machine-learning models - Random Forest, Support Vector Machine, Logistic Regression, XGBoost, and Deep Learning — on features derived from multiple training datasets, optimised to balance specificity, sensitivity, and F1 score. To improve robustness, the two sources are combined through triangulation: an account is classified as a bot only if flagged as such by both Botometer and at least one of the trained models.

Using this methodology, all users active at the protest peak (k-core $\geq$ 2) were classified as `human', `bot,' or `unknown'. A detailed description of the methodology and model performance is provided in that earlier work \cite{li2024socialbotssouractivist}.


\section*{Opinion categorization using ChatGPT}
We used ChatGPT 3.5\cite{chatgpt} to categorise the users' opinion on protests.The efficacy of leveraging large language models (LLMs) for these purposes has been widely explored in academic studies, and the methodology has been employed in prior research addressing similar textual classification tasks\cite{zhu2023can,wu2023event}. 

The full prompt is as follows:

\begin{quote}
On the text that I will give you in the following dialogues, please interpret it as a human with common political sense and background knowledge of racial justice issues. Tell me how positive do you think is this user's opinion towards the BLM protes. Give me a score that falls in the continuous range of -1 to 1. -1 being extremely negative, 1 extremely positive, 0 being neutral.
If you think it is irrelevant to BLM, put 0 there.
If the opinion seem to be mixed, put 0. Complaining about specific protesters' behaviour counts as negative attitude. Arguing that the protest's aim, goal or view of BLM is too extreme counts as a mild negative. Beware of sarcasm. People can support or do not support BLM protests with different partisan preferences, do not take that into account.Just return a score, no explanation needed. 
No full sentence needed either, the number itself will be fine. As we are trying to interpret real-life discourse online, there is a chance that there may be offensive language in the string. It is a virtual experiment evaluating people’s attitude, 
no one is actually hurt during the process. 
Here is the text: 
\end{quote}

We validated the GPT support classification against manual coding using data from a similar online activism event, applying the same prompt instrument used for the Black Lives Matter analysis. We drew a random sample of N=300 users and asked two independent coders to replicate the classification from exactly the same input given to GPT (each user's combined timeline). Table~\ref{chatgpt-pearson} reports Pearson correlations among the GPT-3.5 scores, the two coders, and the coders' averaged score. The two coders correlate at C=0.71, and their averaged judgement correlates with GPT's output at C=0.88
that is, GPT agrees with the human consensus at least as strongly as the coders agree with one another. We therefore treat the GPT support classification as a valid automated measure of user support level.

\begin{table}[ht]
\caption{Pearson Correlation between the support level values of Twitter users labelled by ChatGPT and manual coders.}
\begin{tabular}{lllll}
\toprule
           & ChatGPT & Average score  & coder 1 & coder 2       \\
\midrule
ChatGPT    & 1.00              & 0.88    & 0.72    & 0.91 \\
Average score & 0.88              & 1.00    & 0.92    & 0.93 \\
coder 1    & 0.72              & 0.92    & 1.00    & 0.71 \\
coder 2    & 0.91              & 0.93    & 0.71    & 1.00 \\
\bottomrule
\end{tabular}
\label{chatgpt-pearson}
\end{table}

\section*{Network construction and filtering}
In this study, we constructed a bi-directional, unweighted retweet network (RT network) where nodes represent users and edges represent retweets or quotes between them. Edges were unweighted to reflect the presence of connections, simplifying the analysis by aggregating multiple interactions between node pairs into single edges. This approach helps to explore how interactions with bots might influence connections among human users.

`Retweets' are defined as posts shared with no additional comments, while `quotes' include extra commentary. Each retweet or quote was recorded as a directional edge, with the original tweet poster as the starting node and the retweeter or commenter as the end node. Self-loops, indicating users who retweeted or quoted their own tweets, were included due to their impact on online discourse and mobilisation. Isolated nodes, representing users without original tweets or retweets, were removed.

For the analysis, we focused on a community of users who interacted with bots during a 48-hour snapshot of the BLM communication network. We used the k-core metric to filter the network, focusing on the maximal subgraph where each node is connected to at least k other nodes. We retained users with a k-core value greater than two, resulting in a filtered network of 92,078 users and 372,932 edges. The filtering process was conducted using Python's NetworkX package and Gephi.

To measure the impact of bot interactions on network cohesiveness among human nodes, we constructed two additional network snapshots: the `before' network (May 1, 2020, to May 2, 2020) and the `after' network (September 1, 2020, to September 2, 2020). Directed communication networks were based on retweeting information, with full timeline data and user profiles collected for these periods.

\begin{table}[]
\centering
\footnotesize
\renewcommand{\arraystretch}{1.2}
\caption{Descriptive statistics for k-core2 networks before, during and after protest peak}
\begin{tabular}{@{}lccc@{}}
\toprule
                               & Before   & Peak     & After    \\ \midrule
Number of vertices (N)         & 92078    & 103050   & 92078    \\
Number of edges                & 740742   & 328708   & 913151   \\
Average unweighted degree      & 16.09    & 6.38     & 19.83    \\
Maximum in-degree              & 9155     & 12605    & 10747    \\
Maximum out-degree             & 313      & 77       & 239      \\
Average human clustering coefficient & 0.028 & 0.029 & 0.030 \\
Human--human density           & 0.000073 & 0.000056 & 0.000090 \\
Average bot--human density     & 0.015    & 0.112    & 0.014    \\
Number of active bots          & 39661    & 53327    & 38222    \\
Number of active humans        & 17541    & 32943    & 17255    \\
Number of active communities (CPM) & 1279 & 2831  & 1118     \\ \bottomrule
\end{tabular}
\label{descriptive_stat_network}
\end{table}

In Table \ref{descriptive_stat_network} above, we present descriptive statistics for the network snapshots analysed, including the number of users, number of edges, and the average degree for each snapshot. It is important to note that the differences in the number of edges among the networks are attributable to the sampling methods used. The peak network, specifically, was sampled purely for analytical purposes and only included edges that are directly relevant to the protest. Our main focus is on understanding the changes between the `before' and `after' periods to evaluate the impact of the protests on network structure and bot interactions.

\clearpage

\section*{Model tables}

\begin{table}[h!tbp] \centering
\footnotesize
  \caption{Individual-level OLS models (bot threshold = 0.65)}
  \label{tab:ego_t065}
\begin{tabular}{@{\extracolsep{5pt}}lccccc}
\\[-1.8ex]\hline
\hline \\[-1.8ex]
 & \multicolumn{5}{c}{\textit{Dependent variable:}} \\
\\[-1.8ex] & \multicolumn{5}{c}{Change in local clustering (human-only, scaled)} \\
 & (1) & (2) & (3) & (4) & (5) \\
\\[-1.8ex]\hline \\[-1.8ex]
 Bot interactions (log) & -0.2819 *** & -0.0281 & -0.0401 * & -0.0276 & 0.1539 *** \\
  & (0.0299) & (0.0173) & (0.0178) & (0.0184) & (0.0460) \\
  &  &  &  &  &  \\
 Baseline human clustering &  & -1.4232 *** & -1.4335 *** & -1.4247 *** & -1.4223 *** \\
  &  & (0.0172) & (0.0180) & (0.0183) & (0.0183) \\
  &  &  &  &  &  \\
 Friend--follower ratio (log) &  &  & -0.0349 & -0.0439 * & -0.0423 * \\
  &  &  & (0.0186) & (0.0189) & (0.0189) \\
  &  &  &  &  &  \\
 Account age &  &  & -0.0067 *** & -0.0077 *** & -0.008 *** \\
  &  &  & (0.0016) & (0.0017) & (0.0017) \\
  &  &  &  &  &  \\
 Statuses (log) &  &  & -0.01 & -0.0068 & -0.0067 \\
  &  &  & (0.0056) & (0.0057) & (0.0057) \\
  &  &  &  &  &  \\
 Favourites (log) &  &  & 0.0141 ** & 0.0137 ** & 0.0137 ** \\
  &  &  & (0.0052) & (0.0052) & (0.0052) \\
  &  &  &  &  &  \\
 Pro-BLM opinion (binary) &  &  &  & -0.0365 ** & -0.0171 \\
  &  &  &  & (0.0134) & (0.0141) \\
  &  &  &  &  &  \\
 Bot interactions (log) x Pro-BLM opinion &  &  &  &  & -0.2158 *** \\
  &  &  &  &  & (0.0502) \\
  &  &  &  &  &  \\
 Constant & 0.2765 *** & 0.3929 *** & 0.4001 *** & 0.3989 *** & 0.3885 *** \\
  & (0.0092) & (0.0054) & (0.0081) & (0.0081) & (0.0084) \\
  &  &  &  &  &  \\
\hline \\[-1.8ex]
Observations & 3,291 & 3,291 & 3,291 & 3,291 & 3,291 \\
$R^{2}$ & 0.026 & 0.684 & 0.688 & 0.688 & 0.690 \\
Adjusted $R^{2}$ & 0.026 & 0.684 & 0.687 & 0.688 & 0.689 \\
\hline
\hline \\[-1.8ex]
\textit{Note:}  & \multicolumn{5}{r}{* $p<0.05$; ** $p<0.01$; *** $p<0.001$} \\
\end{tabular}
\end{table}
\clearpage

\begin{table}[tbp] \centering
\footnotesize
  \caption{Individual-level OLS models (bot threshold = 0.7)}
  \label{tab:ego_t07}
\begin{tabular}{@{\extracolsep{5pt}}lccccc}
\\[-1.8ex]\hline
\hline \\[-1.8ex]
 & \multicolumn{5}{c}{\textit{Dependent variable:}} \\
\\[-1.8ex] & \multicolumn{5}{c}{Change in local clustering (human-only, scaled)} \\
 & (1) & (2) & (3) & (4) & (5) \\
\\[-1.8ex]\hline \\[-1.8ex]
 Bot interactions (log) & -0.2816 *** & -0.0291 & -0.041 * & -0.0283 & 0.1569 *** \\
  & (0.0307) & (0.0177) & (0.0182) & (0.0188) & (0.0460) \\
  &  &  &  &  &  \\
 Baseline human clustering &  & -1.4232 *** & -1.4336 *** & -1.4248 *** & -1.4224 *** \\
  &  & (0.0172) & (0.0180) & (0.0183) & (0.0183) \\
  &  &  &  &  &  \\
 Friend--follower ratio (log) &  &  & -0.0349 & -0.044 * & -0.0424 * \\
  &  &  & (0.0186) & (0.0189) & (0.0189) \\
  &  &  &  &  &  \\
 Account age &  &  & -0.0067 *** & -0.0077 *** & -0.008 *** \\
  &  &  & (0.0016) & (0.0017) & (0.0017) \\
  &  &  &  &  &  \\
 Statuses (log) &  &  & -0.0101 & -0.0068 & -0.0068 \\
  &  &  & (0.0056) & (0.0057) & (0.0057) \\
  &  &  &  &  &  \\
 Favourites (log) &  &  & 0.0141 ** & 0.0137 ** & 0.0137 ** \\
  &  &  & (0.0052) & (0.0052) & (0.0052) \\
  &  &  &  &  &  \\
 Pro-BLM opinion (binary) &  &  &  & -0.0366 ** & -0.0167 \\
  &  &  &  & (0.0134) & (0.0141) \\
  &  &  &  &  &  \\
 Bot interactions (log) x Pro-BLM opinion &  &  &  &  & -0.222 *** \\
  &  &  &  &  & (0.0504) \\
  &  &  &  &  &  \\
 Constant & 0.2754 *** & 0.3929 *** & 0.4002 *** & 0.399 *** & 0.3884 *** \\
  & (0.0092) & (0.0054) & (0.0081) & (0.0081) & (0.0084) \\
  &  &  &  &  &  \\
\hline \\[-1.8ex]
Observations & 3,291 & 3,291 & 3,291 & 3,291 & 3,291 \\
$R^{2}$ & 0.025 & 0.684 & 0.688 & 0.688 & 0.690 \\
Adjusted $R^{2}$ & 0.025 & 0.684 & 0.687 & 0.688 & 0.689 \\
\hline
\hline \\[-1.8ex]
\textit{Note:}  & \multicolumn{5}{r}{* $p<0.05$; ** $p<0.01$; *** $p<0.001$} \\
\end{tabular}
\end{table}
\clearpage

\begin{table}[tbp] \centering
\footnotesize
  \caption{Individual-level OLS models (bot threshold = 0.75)}
  \label{tab:ego_t075}
\begin{tabular}{@{\extracolsep{5pt}}lccccc}
\\[-1.8ex]\hline
\hline \\[-1.8ex]
 & \multicolumn{5}{c}{\textit{Dependent variable:}} \\
\\[-1.8ex] & \multicolumn{5}{c}{Change in local clustering (human-only, scaled)} \\
 & (1) & (2) & (3) & (4) & (5) \\
\\[-1.8ex]\hline \\[-1.8ex]
 Bot interactions (log) & -0.2819 *** & -0.032 & -0.0441 * & -0.0316 & 0.1496 ** \\
  & (0.0315) & (0.0182) & (0.0186) & (0.0192) & (0.0466) \\
  &  &  &  &  &  \\
 Baseline human clustering &  & -1.4231 *** & -1.4335 *** & -1.4247 *** & -1.4227 *** \\
  &  & (0.0172) & (0.0180) & (0.0183) & (0.0183) \\
  &  &  &  &  &  \\
 Friend--follower ratio (log) &  &  & -0.0352 & -0.0441 * & -0.042 * \\
  &  &  & (0.0186) & (0.0189) & (0.0189) \\
  &  &  &  &  &  \\
 Account age &  &  & -0.0067 *** & -0.0077 *** & -0.008 *** \\
  &  &  & (0.0016) & (0.0017) & (0.0017) \\
  &  &  &  &  &  \\
 Statuses (log) &  &  & -0.0101 & -0.0068 & -0.0066 \\
  &  &  & (0.0056) & (0.0057) & (0.0057) \\
  &  &  &  &  &  \\
 Favourites (log) &  &  & 0.0142 ** & 0.0137 ** & 0.0136 ** \\
  &  &  & (0.0052) & (0.0052) & (0.0052) \\
  &  &  &  &  &  \\
 Pro-BLM opinion (binary) &  &  &  & -0.0363 ** & -0.0176 \\
  &  &  &  & (0.0134) & (0.0140) \\
  &  &  &  &  &  \\
 Bot interactions (log) x Pro-BLM opinion &  &  &  &  & -0.218 *** \\
  &  &  &  &  & (0.0511) \\
  &  &  &  &  &  \\
 Constant & 0.2737 *** & 0.3931 *** & 0.4003 *** & 0.3992 *** & 0.3888 *** \\
  & (0.0092) & (0.0054) & (0.0081) & (0.0081) & (0.0084) \\
  &  &  &  &  &  \\
\hline \\[-1.8ex]
Observations & 3,291 & 3,291 & 3,291 & 3,291 & 3,291 \\
$R^{2}$ & 0.024 & 0.684 & 0.688 & 0.688 & 0.690 \\
Adjusted $R^{2}$ & 0.023 & 0.684 & 0.687 & 0.688 & 0.689 \\
\hline
\hline \\[-1.8ex]
\textit{Note:}  & \multicolumn{5}{r}{* $p<0.05$; ** $p<0.01$; *** $p<0.001$} \\
\end{tabular}
\end{table}
\clearpage

\begin{table}[tbp] \centering
\footnotesize
  \caption{Community-level OLS models (bot threshold = 0.65)}
  \label{tab:comm_t065}
\begin{tabular}{@{\extracolsep{5pt}}lccccc}
\\[-1.8ex]\hline
\hline \\[-1.8ex]
 & \multicolumn{5}{c}{\textit{Dependent variable:}} \\
\\[-1.8ex] & \multicolumn{5}{c}{Change in human network density during BLM} \\
 & (1) & (2) & (3) & (4) & (5) \\
\\[-1.8ex]\hline \\[-1.8ex]
 Log bot--human density & -0.871 *** & -0.8936 *** & -0.7052 *** & -0.6905 *** & -0.4426 \\
  & (0.1535) & (0.1642) & (0.1835) & (0.1834) & (0.2540) \\
  &  &  &  &  &  \\
 Baseline human density &  & 0.0179 & 0.0365 & 0.0293 & 0.0276 \\
  &  & (0.0461) & (0.0527) & (0.0528) & (0.0527) \\
  &  &  &  &  &  \\
 Log number of humans &  &  & 0.0509 & 0.0538 & 0.0509 \\
  &  &  & (0.0274) & (0.0274) & (0.0275) \\
  &  &  &  &  &  \\
 Median statuses (log) &  &  & 0.0185 * & 0.0189 * & 0.0196 * \\
  &  &  & (0.0087) & (0.0087) & (0.0087) \\
  &  &  &  &  &  \\
 Median favourites (log) &  &  & 0.0066 & 0.0066 & 0.0068 \\
  &  &  & (0.0081) & (0.0081) & (0.0081) \\
  &  &  &  &  &  \\
 Friend--follower ratio &  &  & 0.0012 & 0.0013 & 0.0012 \\
  &  &  & (0.0007) & (0.0007) & (0.0007) \\
  &  &  &  &  &  \\
 Opinion during peak &  &  &  & -0.0193 & 0.0024 \\
  &  &  &  & (0.0106) & (0.0187) \\
  &  &  &  &  &  \\
 Log density x opinion &  &  &  &  & -0.4388 \\
  &  &  &  &  & (0.3114) \\
  &  &  &  &  &  \\
 Constant & -0.0625 *** & -0.0629 *** & -0.3896 *** & -0.3838 *** & -0.403 *** \\
  & (0.0090) & (0.0091) & (0.1029) & (0.1028) & (0.1036) \\
  &  &  &  &  &  \\
\hline \\[-1.8ex]
Observations & 642 & 642 & 642 & 642 & 642 \\
$R^{2}$ & 0.048 & 0.048 & 0.068 & 0.073 & 0.076 \\
Adjusted $R^{2}$ & 0.046 & 0.045 & 0.059 & 0.063 & 0.064 \\
\hline
\hline \\[-1.8ex]
\textit{Note:}  & \multicolumn{5}{r}{* $p<0.05$; ** $p<0.01$; *** $p<0.001$} \\
\end{tabular}
\end{table}
\clearpage

\begin{table}[tbp] \centering
\footnotesize
  \caption{Community-level OLS models (bot threshold = 0.7)}
  \label{tab:comm_t07}
\begin{tabular}{@{\extracolsep{5pt}}lccccc}
\\[-1.8ex]\hline
\hline \\[-1.8ex]
 & \multicolumn{5}{c}{\textit{Dependent variable:}} \\
\\[-1.8ex] & \multicolumn{5}{c}{Change in human network density during BLM} \\
 & (1) & (2) & (3) & (4) & (5) \\
\\[-1.8ex]\hline \\[-1.8ex]
 Log bot--human density & -0.6352 *** & -0.6278 *** & -0.4169 * & -0.4075 * & -0.2242 \\
  & (0.1454) & (0.1533) & (0.1662) & (0.1661) & (0.2529) \\
  &  &  &  &  &  \\
 Baseline human density &  & -0.007 & 0.0314 & 0.0242 & 0.0219 \\
  &  & (0.0459) & (0.0530) & (0.0531) & (0.0532) \\
  &  &  &  &  &  \\
 Log number of humans &  &  & 0.0687 * & 0.0712 ** & 0.071 ** \\
  &  &  & (0.0268) & (0.0268) & (0.0268) \\
  &  &  &  &  &  \\
 Median statuses (log) &  &  & 0.0192 * & 0.0196 * & 0.0204 * \\
  &  &  & (0.0088) & (0.0088) & (0.0088) \\
  &  &  &  &  &  \\
 Median favourites (log) &  &  & 0.0082 & 0.0081 & 0.0081 \\
  &  &  & (0.0082) & (0.0082) & (0.0082) \\
  &  &  &  &  &  \\
 Friend--follower ratio &  &  & 0.0013 & 0.0013 & 0.0013 \\
  &  &  & (0.0007) & (0.0007) & (0.0007) \\
  &  &  &  &  &  \\
 Opinion during peak &  &  &  & -0.0185 & -0.0044 \\
  &  &  &  & (0.0107) & (0.0182) \\
  &  &  &  &  &  \\
 Log density x opinion &  &  &  &  & -0.2881 \\
  &  &  &  &  & (0.2999) \\
  &  &  &  &  &  \\
 Constant & -0.0728 *** & -0.0725 *** & -0.4435 *** & -0.4366 *** & -0.4538 *** \\
  & (0.0087) & (0.0088) & (0.1033) & (0.1033) & (0.1048) \\
  &  &  &  &  &  \\
\hline \\[-1.8ex]
Observations & 632 & 632 & 632 & 632 & 632 \\
$R^{2}$ & 0.029 & 0.029 & 0.057 & 0.061 & 0.062 \\
Adjusted $R^{2}$ & 0.028 & 0.026 & 0.047 & 0.050 & 0.050 \\
\hline
\hline \\[-1.8ex]
\textit{Note:}  & \multicolumn{5}{r}{* $p<0.05$; ** $p<0.01$; *** $p<0.001$} \\
\end{tabular}
\end{table}
\clearpage

\begin{table}[tbp] \centering
\footnotesize
  \caption{Community-level OLS models (bot threshold = 0.75)}
  \label{tab:comm_t075}
\begin{tabular}{@{\extracolsep{5pt}}lccccc}
\\[-1.8ex]\hline
\hline \\[-1.8ex]
 & \multicolumn{5}{c}{\textit{Dependent variable:}} \\
\\[-1.8ex] & \multicolumn{5}{c}{Change in human network density during BLM} \\
 & (1) & (2) & (3) & (4) & (5) \\
\\[-1.8ex]\hline \\[-1.8ex]
 Log bot--human density & -0.5714 *** & -0.5944 *** & -0.414 ** & -0.4025 * & -0.2515 \\
  & (0.1418) & (0.1486) & (0.1586) & (0.1587) & (0.2583) \\
  &  &  &  &  &  \\
 Baseline human density &  & 0.0236 & 0.0663 & 0.0605 & 0.0583 \\
  &  & (0.0454) & (0.0526) & (0.0528) & (0.0529) \\
  &  &  &  &  &  \\
 Log number of humans &  &  & 0.0693 ** & 0.0711 ** & 0.0709 ** \\
  &  &  & (0.0262) & (0.0262) & (0.0262) \\
  &  &  &  &  &  \\
 Median statuses (log) &  &  & 0.0225 * & 0.023 ** & 0.0235 ** \\
  &  &  & (0.0089) & (0.0089) & (0.0089) \\
  &  &  &  &  &  \\
 Median favourites (log) &  &  & -0.001 & -0.0011 & -0.0011 \\
  &  &  & (0.0083) & (0.0083) & (0.0083) \\
  &  &  &  &  &  \\
 Friend--follower ratio &  &  & 0.0014 & 0.0014 * & 0.0014 * \\
  &  &  & (0.0007) & (0.0007) & (0.0007) \\
  &  &  &  &  &  \\
 Opinion during peak &  &  &  & -0.0151 & -0.0045 \\
  &  &  &  & (0.0107) & (0.0179) \\
  &  &  &  &  &  \\
 Log density x opinion &  &  &  &  & -0.2225 \\
  &  &  &  &  & (0.3002) \\
  &  &  &  &  &  \\
 Constant & -0.0748 *** & -0.0757 *** & -0.3828 *** & -0.3784 *** & -0.39 *** \\
  & (0.0084) & (0.0086) & (0.1024) & (0.1024) & (0.1036) \\
  &  &  &  &  &  \\
\hline \\[-1.8ex]
Observations & 602 & 602 & 602 & 602 & 602 \\
$R^{2}$ & 0.026 & 0.027 & 0.054 & 0.057 & 0.058 \\
Adjusted $R^{2}$ & 0.025 & 0.024 & 0.045 & 0.046 & 0.046 \\
\hline
\hline \\[-1.8ex]
\textit{Note:}  & \multicolumn{5}{r}{* $p<0.05$; ** $p<0.01$; *** $p<0.001$} \\
\end{tabular}
\end{table}
\clearpage

\begin{table}[tbp] \centering
\footnotesize
  \caption{Robustness: individual models with BLM-keyword-filtered after period (bot threshold = 0.65). Columns (3)/(5) reuse the main Model 3/5 specifications; ``Full after'' uses the unrestricted after timeline and ``BLM after'' recomputes after-period outcomes from BLM-related retweets only (peak exposure unchanged).}
  \label{tab:ego_blm_after_robust_t065}
\begin{tabular}{@{\extracolsep{5pt}}lcccc}
\\[-1.8ex]\hline
\hline \\[-1.8ex]
 & \multicolumn{4}{c}{\textit{Dependent variable:}} \\
\\[-1.8ex] & \multicolumn{4}{c}{Change in local clustering (human-only, scaled)} \\
 & (3) Full after & (3) BLM after & (5) Full after & (5) BLM after \\
\\[-1.8ex]\hline \\[-1.8ex]
 Bot interactions (log) & -0.0401 * & -0.0237 *** & 0.1539 *** & 0.0177 \\
  & (0.0178) & (0.0061) & (0.0460) & (0.0432) \\
  &  &  &  &  \\
 Baseline human clustering & -1.4335 *** & -0.7659 *** & -1.4223 *** & -0.7656 *** \\
  & (0.0180) & (0.0065) & (0.0183) & (0.0065) \\
  &  &  &  &  \\
 Friend--follower ratio (log) & -0.0349 & -0.0137 & -0.0423 * & -0.0173 * \\
  & (0.0186) & (0.0079) & (0.0189) & (0.0080) \\
  &  &  &  &  \\
 Account age & -0.0067 *** & 0.0027 *** & -0.008 *** & 0.0022 ** \\
  & (0.0016) & (0.0008) & (0.0017) & (0.0008) \\
  &  &  &  &  \\
 Statuses (log) & -0.01 & -0.0106 *** & -0.0067 & -0.0093 ** \\
  & (0.0056) & (0.0031) & (0.0057) & (0.0031) \\
  &  &  &  &  \\
 Favourites (log) & 0.0141 ** & -0.0002 & 0.0137 ** & 0.0003 \\
  & (0.0052) & (0.0029) & (0.0052) & (0.0029) \\
  &  &  &  &  \\
 Pro-BLM opinion (binary) &  &  & -0.0171 & -0.0186 * \\
  &  &  & (0.0141) & (0.0084) \\
  &  &  &  &  \\
 Bot interactions (log) x Pro-BLM opinion &  &  & -0.2158 *** & -0.0381 \\
  &  &  & (0.0502) & (0.0436) \\
  &  &  &  &  \\
 Constant & 0.4001 *** & -0.15 *** & 0.3885 *** & -0.1514 *** \\
  & (0.0081) & (0.0093) & (0.0084) & (0.0093) \\
  &  &  &  &  \\
\hline \\[-1.8ex]
Observations & 3,291 & 714 & 3,291 & 714 \\
$R^{2}$ & 0.688 & 0.961 & 0.690 & 0.961 \\
Adjusted $R^{2}$ & 0.687 & 0.960 & 0.689 & 0.961 \\
\hline
\hline \\[-1.8ex]
\textit{Note:}  & \multicolumn{4}{r}{* $p<0.05$; ** $p<0.01$; *** $p<0.001$} \\
\end{tabular}
\end{table}
\clearpage

\begin{table}[tbp] \centering
\footnotesize
  \caption{Robustness: community models with BLM-keyword-filtered after period (bot threshold = 0.65). Columns (3)/(5) reuse the main Model 3/5 specifications; ``Full after'' uses the unrestricted after timeline and ``BLM after'' recomputes after-period outcomes from BLM-related retweets only (peak exposure unchanged).}
  \label{tab:comm_blm_after_robust_t065}
\begin{tabular}{@{\extracolsep{5pt}}lcccc}
\\[-1.8ex]\hline
\hline \\[-1.8ex]
 & \multicolumn{4}{c}{\textit{Dependent variable:}} \\
\\[-1.8ex] & \multicolumn{4}{c}{Change in human network density during BLM} \\
 & (3) Full after & (3) BLM after & (5) Full after & (5) BLM after \\
\\[-1.8ex]\hline \\[-1.8ex]
 Log bot--human density & -0.7052 *** & -1.0183 *** & -0.4426 & -0.7148 * \\
  & (0.1835) & (0.2207) & (0.2540) & (0.2956) \\
  &  &  &  &  \\
 Baseline human density & 0.0365 & -0.3432 *** & 0.0276 & -0.3623 *** \\
  & (0.0527) & (0.0674) & (0.0527) & (0.0666) \\
  &  &  &  &  \\
 Log number of humans & 0.0509 & 0.1455 *** & 0.0509 & 0.1469 *** \\
  & (0.0274) & (0.0368) & (0.0275) & (0.0364) \\
  &  &  &  &  \\
 Median statuses (log) & 0.0185 * & -0.0017 & 0.0196 * & 0.0022 \\
  & (0.0087) & (0.0105) & (0.0087) & (0.0104) \\
  &  &  &  &  \\
 Median favourites (log) & 0.0066 & -0.0039 & 0.0068 & -0.004 \\
  & (0.0081) & (0.0102) & (0.0081) & (0.0100) \\
  &  &  &  &  \\
 Friend--follower ratio & 0.0012 & 0.0007 & 0.0012 & 0.0008 \\
  & (0.0007) & (0.0009) & (0.0007) & (0.0009) \\
  &  &  &  &  \\
 Opinion during peak &  &  & 0.0024 & -0.0211 \\
  &  &  & (0.0187) & (0.0222) \\
  &  &  &  &  \\
 Log density x opinion &  &  & -0.4388 & -0.4989 \\
  &  &  & (0.3114) & (0.3710) \\
  &  &  &  &  \\
 Constant & -0.3896 *** & -0.1693 & -0.403 *** & -0.1928 \\
  & (0.1029) & (0.1266) & (0.1036) & (0.1262) \\
  &  &  &  &  \\
\hline \\[-1.8ex]
Observations & 642 & 462 & 642 & 462 \\
$R^{2}$ & 0.068 & 0.356 & 0.076 & 0.377 \\
Adjusted $R^{2}$ & 0.059 & 0.347 & 0.064 & 0.366 \\
\hline
\hline \\[-1.8ex]
\textit{Note:}  & \multicolumn{4}{r}{* $p<0.05$; ** $p<0.01$; *** $p<0.001$} \\
\end{tabular}
\end{table}
\clearpage

\begin{table}[!htbp] \centering
  \caption{Ego-level network autocorrelation robustness (bot threshold = 0.65)}
  \label{tab:ego_network_error_t065}
  \resizebox{\textwidth}{!}{%
  \begin{tabular}{@{\extracolsep{5pt}}lcccccc}
  \\[-1.8ex]\hline
  \hline \\[-1.8ex]
   & \multicolumn{3}{c}{Model 3} & \multicolumn{3}{c}{Model 5} \\
  \cline{2-4}\cline{5-7}
   & OLS & Network error & Network error & OLS & Network error & Network error \\
   & None & Ego overlap & CPM community & None & Ego overlap & CPM community \\
  \hline \\[-1.8ex]
  Constant & 0.4001 *** & 0.4080 *** & 0.3991 *** & 0.3885 *** & 0.3982 *** & 0.3879 *** \\
   & (0.0081) & (0.0080) & (0.0082) & (0.0084) & (0.0083) & (0.0085) \\
   & & & & & & \\
  Baseline human clustering & -1.4335 *** & -1.3934 *** & -1.4265 *** & -1.4223 *** & -1.3836 *** & -1.4156 *** \\
   & (0.0180) & (0.0203) & (0.0181) & (0.0183) & (0.0204) & (0.0183) \\
   & & & & & & \\
  Bot interactions (log) & -0.0401 * & -0.0404 * & -0.0434 * & 0.1539 *** & 0.1253 ** & 0.1509 ** \\
   & (0.0178) & (0.0194) & (0.0180) & (0.0460) & (0.0451) & (0.0461) \\
   & & & & & & \\
  Friend--follower ratio (log) & -0.0349 & -0.0337 & -0.0351 & -0.0423 * & -0.0415 * & -0.0425 * \\
   & (0.0186) & (0.0209) & (0.0189) & (0.0189) & (0.0210) & (0.0191) \\
   & & & & & & \\
  Account age & -0.0067 *** & -0.0030 & -0.0063 *** & -0.0080 *** & -0.0045 * & -0.0075 *** \\
   & (0.0016) & (0.0020) & (0.0016) & (0.0017) & (0.0020) & (0.0017) \\
   & & & & & & \\
  Statuses (log) & -0.0100 & -0.0112 & -0.0088 & -0.0067 & -0.0079 & -0.0055 \\
   & (0.0056) & (0.0064) & (0.0056) & (0.0057) & (0.0065) & (0.0057) \\
   & & & & & & \\
  Favourites (log) & 0.0141 ** & 0.0127 * & 0.0128 * & 0.0137 ** & 0.0127 * & 0.0124 * \\
   & (0.0052) & (0.0060) & (0.0052) & (0.0052) & (0.0059) & (0.0052) \\
   & & & & & & \\
  Pro-BLM opinion (binary) &  &  &  & -0.0171 & -0.0262 & -0.0167 \\
   &  &  &  & (0.0141) & (0.0158) & (0.0142) \\
   & & & & & & \\
  Bot interactions (log) x Pro-BLM opinion &  &  &  & -0.2158 *** & -0.1905 *** & -0.2169 *** \\
   &  &  &  & (0.0502) & (0.0500) & (0.0503) \\
   & & & & & & \\
  Network error parameter ($\lambda$) &  & 0.3414 *** & 0.0753 *** &  & 0.3377 *** & 0.0750 *** \\
   &  & (0.0314) & (0.0201) &  & (0.0315) & (0.0201) \\
   & & & & & & \\
  \hline \\[-1.8ex]
  Observations & 3291 & 3291 & 3291 & 3291 & 3291 & 3291 \\
  $R^2$/pseudo-$R^2$ & 0.6876 & 0.6870 & 0.6876 & 0.6900 & 0.6894 & 0.6900 \\
  Filtered residual Moran's $I$ &  & -0.0204 & -0.0024 &  & -0.0192 & -0.0022 \\
  Moran permutation $p$ &  & 0.0780 & 0.9330 &  & 0.1130 & 0.9500 \\
  \hline
  \hline \\[-1.8ex]
  \end{tabular}}
  \begin{minipage}{0.98\textwidth}\footnotesize \textit{Note:} Maximum-likelihood network error models use row-standardized weights. Ego overlap is Jaccard similarity between closed human ego neighbourhoods, retaining the top 20 neighbours per ego before symmetrization. CPM weights connect observations assigned to the same peak community. Zero-neighbour observations are retained as islands. Moran tests use 999 permutations. * $p<0.05$; ** $p<0.01$; *** $p<0.001$.\end{minipage}
\end{table}



\end{document}